\documentclass[10pt, conference, letterpaper]{IEEEtran}
\IEEEoverridecommandlockouts
\usepackage[T1]{fontenc}
\usepackage[english]{babel}
\usepackage[utf8]{inputenc}
\usepackage{comment,acronym,amsthm,amsmath,amssymb,tabularx,bbm,algorithmic,subcaption,epsfig,xspace,multirow}
\usepackage[algoruled,vlined,linesnumbered]{algorithm2e}
\usepackage[dvipsnames]{xcolor}
\usepackage[font={small}]{caption}
\usepackage{rotating}
\usepackage{flushend}
\usepackage{times}
\usepackage{url}

\newtheorem{prop}{Proposition}

\def \N{{\mathcal N}}
\def \K{{\mathcal K}}

\def \U{{\mathcal U}}

\newcommand{\Servers}[1]{\ensuremath{S_{#1}}\xspace}
\newcommand{\Apps}[1]{\ensuremath{U_{#1}}\xspace}

\newcommand{\BandR}[2]{\ensuremath{B_{{#1}{#2}}}\xspace}
\newcommand{\BandA}[1]{\ensuremath{B_{#1}}\xspace}
\newcommand{\DelayR}[2]{\ensuremath{d_{{#1}{#2}}}\xspace} 
\newcommand{\DelayA}[1]{\ensuremath{d_{#1}}\xspace} 
\newcommand{\Unit}[2]{\ensuremath{s_{{#1}_{#2}}}\xspace}
\newcommand{\ThroughL}[1]{\ensuremath{\lambda_{#1}^L}\xspace}
\newcommand{\ThroughH}[1]{\ensuremath{\lambda_{#1}^H}\xspace}
\newcommand{\ReqApp}[1]{\ensuremath{\mathbf{c_{#1}}}\xspace}
\newcommand{\UnitV}[2]{\ensuremath{\mathbf{C_{#1_{#2}}}}\xspace}
\newcommand{\Freq}[1]{\ensuremath{F_{#1}}\xspace}
\newcommand{\Data}[1]{\ensuremath{\Delta_{#1}}\xspace}
\newcommand{\DataL}[1]{\ensuremath{\Delta_{#1}^L}}
\newcommand{\DataH}[1]{\ensuremath{\Delta_{#1}^H}}
\newcommand{\MemApp}[1]{\ensuremath{c^M_{#1}}\xspace}

\newcommand{\ProcApp}[1]{\ensuremath{c^P_{#1}}\xspace}

\newcommand{\StorApp}[1]{\ensuremath{c^S_{#1}}\xspace}

\newcommand{\MemUnit}[2]{\ensuremath{C^M_{{#1}_{#2}}}\xspace}

\newcommand{\ProcUnit}[2]{\ensuremath{C^P_{{#1}_{#2}}}\xspace}

\newcommand{\StorUnit}[2]{\ensuremath{C^S_{{#1}_{#2}}}\xspace}

\newcommand{\ModA}[1]{\ensuremath{{#1}_A}\xspace}
\newcommand{\ModB}[1]{\ensuremath{{#1}_B}\xspace}

\DeclareMathOperator*{\argmin}{arg\,min}

\SetKwInput{Input}{Input}
\SetKwInOut{Output}{Output}
 
\begin{document}
\title{Cutting Throughput on the Edge: \\ App-Aware Placement in Fog Computing}
\author{Francescomaria Faticanti, Francesco De Pellegrini, Domenico Siracusa, Daniele Santoro and Silvio Cretti$^\diamond$\thanks{$^\diamond$Fondazione Bruno Kessler, via Sommarive, 18 I-38123 Povo, Trento, Italy}}
\maketitle
\begin{abstract}
Fog computing extends cloud computing technology to the edge of the infrastructure to let
 IoT applications access objects' data with reduced latency, location awareness and dynamic computation. 
By displacing workloads from the central cloud to the edge devices, fog computing overcomes
communication bottlenecks avoiding raw data transfer to the central cloud, thus paving the way for 
the next generation IoT-based applications.

In this paper we study scheduling and placement of applications in fog computing, which is
key to ensure profitability for the involved stakeholders. We consider a scenario where the
emerging  microservice architecture allows for the design of applications as cascades of coupled microservice
modules. It results into a mixed integer non linear problem involving constraints on both application
data flows and computation placement. Due to the complexity of the original problem, we resort to a
simplified version, which is further solved using a greedy algorithm. This algorithm is the core placement
logic of the FogAtlas platform, a fog computing platform based on existing virtualization technologies.

Extensive numerical results validate the model and the scalability of the proposed solution, showing
it attains performance close to the optimal solution and, in our real implementation, it scales well
with respect to the number of served applications.
\end{abstract}
 

\section{Introduction}


Fog computing adopts cloud technology to move computation to the edge. It promises to solve the core problem
of data explosion in the IoT domain~\cite{IoTSurvey}. Instead of performing raw data transfer to the cloud,
in fact, data flows generated from objects can be intercepted to extract information at the edge of network.
This architectural choice prevents massive, diffused and continuous raw data injection which would ultimately create
severe communication congestion~\cite{bonomi2012fog}. Furthermore, compared to customary cloud-based
IoT deployments, proximity to mobile or sensing devices lowers round-trip-time between objects and backends
of processing applications \cite{Chiang2016}.

Further incentive in the development of fog computing solutions include the standardization of IoT deployments,
ease of management and maintenance of IoT services in industrial networks \cite{Li2018}, and also overcoming privacy
issues by confining raw data within specific geographical regions \cite{Guan2018}. The fog system studied
in this paper refers to FogAtlas, a platform designed to perform efficient deployment of fog computing
applications according to the above guidelines. 

The tradeoff in this context is represented by edge resource occupation: compared to standard cloud
technologies -- based on overprovisioned datacenters -- the business of edge infrastructure
owners will not be able to rely on overprovisioning. Rather, they need to trade off premier
service provision based on localized data processing and low round-trip time for storage,
memory and processing capabilities of edge units \cite{Yu2018Prov,taneja2017resource}.

The paradigm of fog computing consists of a layered architecture, including a central cloud,
a series of edge units, gateways to connect objects and, finally, objects which generate data
and possibly actuate. Virtual machines or containers can run either in the central cloud, or
over edge units, depending on the requirements of IoT-based applications.

To this respect, it is natural to assume that fog-native applications will adhere to the microservice paradigm \cite{Gan2018}.
Microservice applications, in fact, are made by the composition of multiple coupled modules, such as, e.g.,
a graphical user interface, a user repository, a web server, an image recognition module, a monitoring application,
etc. Once interconnected using a specific communication and computing pattern, the microservice architecture
delivers the intended functionality while preserving scalability, minimality and cohesiveness of the
application. In fog computing, the modular structure is indeed appealing in order to simplify the dispatch of
computing modules onto edge nodes.

Typically, the microservice components of an application can be deployed using independent containers.
However, in this work, we make the baseline assumption that a fog application will be shipped using
two modules.  The rationale for such a minimal containerization is that all operations of monitoring and
networking on the edge will be largely simplified. The first one -- possibly a virtual machine hosting
several containers in the cloud -- will typically comprise microservice modules not involved in raw IoT
data computation and can be hosted in the central cloud.

The second module, hosted on a single IoT container, comprises functionalities involving objects'
data processing. This container may reside either on the cloud or on edge nodes, depending
on the scheduling operated by the fog orchestrator. We refer to the concrete example of application
deployed on our FogAtlas platform, namely a plate recognition video application, able to be dispatched on
an edge server close to a target video-camera. Stream mining is actually emerging as a core research
field motivating fog-computing applications \cite{Canzian2015}.

In such benchmark fog application, indeed, performing computing IoT operations directly on edge nodes
provides a clear advantage in terms of bandwidth utilization. In fact, the raw video stream is filtered
through an image detection algorithm so that only tagged frames need to be forwarded to the cloud. As
 a result, only a small fraction of information is transferred toward the central cloud. 

The main objective of this work is to describe an efficient placement of fog applications'
modules either on the edge or in the cloud. In order to determine such a placement,
constraints on computational and bandwidth requirements have to be factored in. We shall introduce
first the general problem  of how to place a batch of applications with sufficient computational
resources and yet efficient network usage. Then, we shall describe our algorithmic solution. 

The rest of the paper is organized as follows. In Sec.~\ref{sec:Kframes} we describe the system model,
including the abstractions we use for the applications' architecture, the network infrastructure and applications' 
deployment configurations. In Sec.~\ref{sec:probform} we present the problem formulation, introducing
the most general problem setting. The placement problem is addressed in Sec.~\ref{sec:place}
by reduction to a multi-dimensional knapsack problem, which can be solved using a greedy algorithm.
The FogAtlas platform is described in~\ref{sec:fogatlas} and numerical results are reported in Sec.~\ref{sec:numres}.
A concluding section ends the paper.


\section{System Model}\label{sec:Kframes}


\begin{table}[t]\caption{Main notation used throughout the paper}
\centering
\begin{tabular}{|p{0.22\columnwidth}|p{0.68\columnwidth}|}
\hline
{\it Symbol} & {\it Meaning}\\
\hline
$\K$ & set of regions $|\K|=K$\\
$\U$ & set of applications to be deployed $\U = \cup_{i=1}^K U_i$, $|\U| = U$\\
\Servers{k} & set of server units in region $k$, with $|\Servers{i}|=n_i, \forall i \in \K$, $\Servers{i} = \{\Unit{i}{1},\dotsc, \Unit{i}{n_i}\}	$\\
\Servers{0} & central cloud\\
\Apps{k} & set of applications requiring IoT data in region $k$\\
\ThroughH{u}/\ThroughL{u} & high/low throughput required by application $u$\\
\DataH{u}/\DataL{u} & large/small data unit of application $u$\\
\Freq{u} & output samples per second required by application $u$\\
$\UnitV{k}{i}$ & memory, storage and processing capacity of the $i$-th server in region $k$: $\UnitV{k}{i}=(\MemUnit{k}{i},\StorUnit{k}{i},\ProcUnit{k}{i})$\\
$\ReqApp{u}$ & memory, storage and processing requirements of application $u$: $\ReqApp{u}=(\MemApp{u},\StorApp{u},\ProcApp{u})$\\
$x_{u,k,i} \in \{0,1\}$, & boolean variable indicating $u$ is placed on server unit $i$ of region $k$\\
 $x_{u,k}$&$x_{u,k} = \sum_{i \in \Servers{k}} x_{u,k,i}$\\
\hline
\end{tabular}\label{tab:notation}
\end{table}

We consider a fog system deployed over a set of geographic regions $\K=\{1,\ldots,K\}$. Region $k$ hosts a set \Servers{k} of edge servers or units. We denote \Unit{k}{i}, with $i \in \{1,\ldots,n_k\}$ a specific edge unit deployed within the $k$-th region; for the sake of notation we denote the
central cloud as \Servers{0}. The resources of edge unit \Unit{k}{i} are represented by capacity vector ${\mathbf C_{k_i}}=(\MemUnit{k}{i},\ProcUnit{k}{i},\StorUnit{k}{i})$. The first component of the capacity vector is the memory capacity. The second component is the processing capacity, which determines the maximum load which
can be sustained on the edge unit. Finally, the third component denotes the storage capacity, i.e., the data  volume that can be accommodated on the storage
of the edge unit. We assume that the storage of a containerized application is handled on the same unit where the container is deployed, with the aim to reduce
the communication costs.

In region $k$, IoT devices serve data required by a set of applications $\Apps{k}$. From here on out, we identify the application and the device from which
data are requested with same symbol. The extension of the following optimization framework in the case of multiple requests for same IoT device is immediate,
by considering virtual replicas of a tagged IoT device. We say that application $u$ ``belongs'' to a given region because the IoT object is located there. Such
region is denoted \Servers{u} for the sake of notation. We leave access of apps to IoT objects of different regions for future works. 

{\noindent \em Network Architecture.} The fog system can be described by a weighted graph $G = (V,E)$ where $V = \{\Servers{i} \cup \Apps{i}\}_{i \in \K}$ and $E \subseteq \binom{V}{2}$. The weight of each edge $(i,j) \in E$ consists of the delay, \DelayR{i}{j}, of the link and the bandwidth of the link \BandR{i}{j}. Let $\N(\Servers{i}) = \{\Servers{j} | (j,i) \in E \}$. 
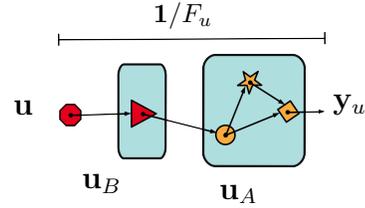
\begin{figure}[t]
	\centering
	\includegraphics[width=0.4\columnwidth]{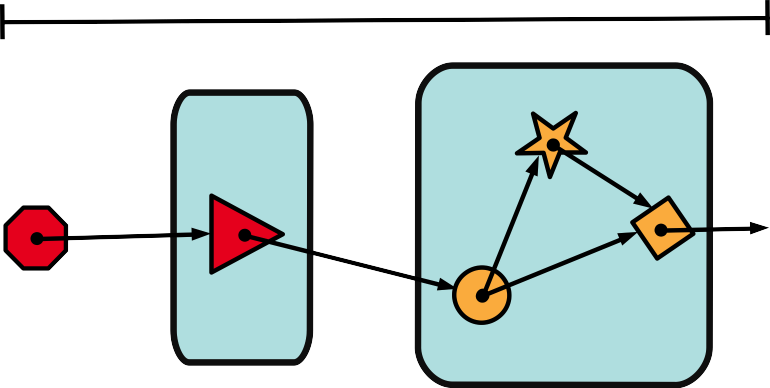}
        \put(-118,20){\large  $\mathbf u$} \put(-92,-8){\large $\mathbf u_B$} \put(-40,-12){\large $\mathbf u_A$}
        \put(3,22){\large $\mathbf y_u$}\put(-65,55){$\mathbf 1/F_u$}
	\caption{The modules cascade outputs a result $y_u$ every $1/F_u$ sec.}
	\label{fig:modules}
\end{figure}

\begin{figure*}[h]
        \centering
        \includegraphics[width=0.23\linewidth]{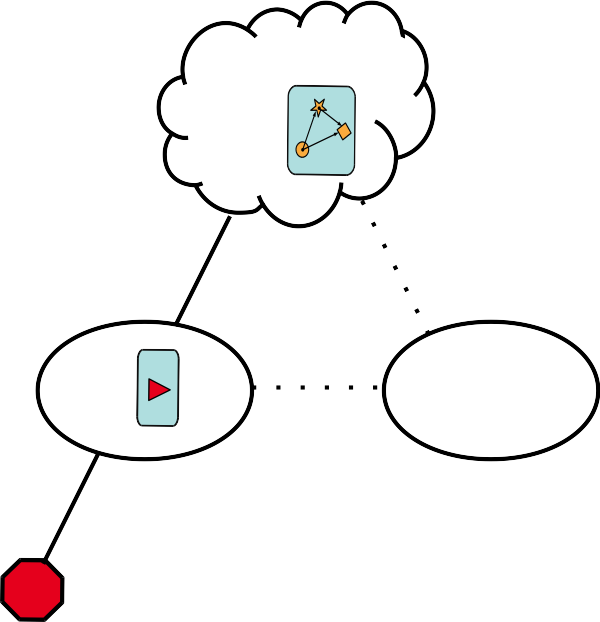}
        \put(0,0){\put(-104,0){\Large $u$}\put(-125,57){\Large $\Servers{u}$}
          \put(-36,118){\Large $\Servers{0}$}\put(-3,57){\Large $\Servers{j}$}
          \put(-110,43){\Large $\ModB{u}$}\put(-64,110){\Large $\ModA{u}$}
          \put(-78,63){$\DataL{u}$, $\ThroughL{u}$}\put(-135,120){\large \underline{Type $1$}}
          \put(-102,19){$\DataH{u}$, $\ThroughH{u}$}
        }
        \centering
        \hskip12mm
        \includegraphics[width=0.23\linewidth]{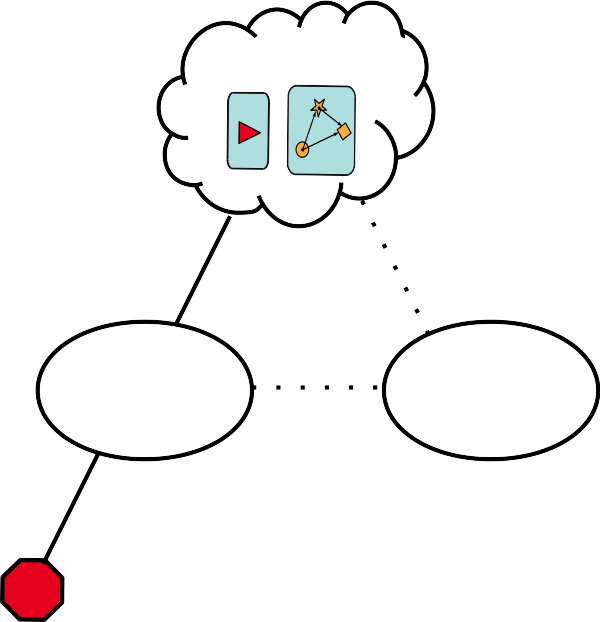}
         \put(0,0){\put(-104,0){\Large $u$}\put(-125,57){\Large $\Servers{u}$}
          \put(-36,118){\Large $\Servers{0}$}\put(-3,57){\Large $\Servers{j}$}
          \put(-79,110){\Large $\ModB{u}$}\put(-61,110){\Large $\ModA{u}$}
          \put(-77,66){$\DataH{u}$, $\ThroughH{u}$}\put(-135,120){\large \underline{Type $2$}}
          \put(-102,19){$\DataH{u}$, $\ThroughH{u}$}
        }
        \centering
        \hskip12mm
        \includegraphics[width=0.23\linewidth]{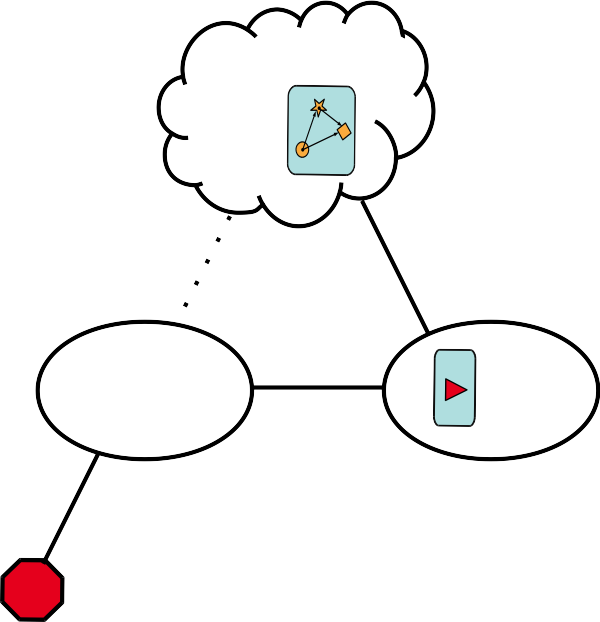}
        \put(0,0){\put(-104,0){\Large $u$}\put(-125,57){\Large $\Servers{u}$}
          \put(-36,118){\Large $\Servers{0}$}\put(-3,57){\Large $\Servers{j}$}
          \put(-20,43){\Large $\ModB{u}$}\put(-64,110){\Large $\ModA{u}$}
          \put(-70,37){$\DataH{u}$,$\ThroughH{u}$}\put(-38,68){$\DataL{u}$, $\ThroughL{u}$}
          \put(-102,19){$\DataH{u}$, $\ThroughH{u}$}\put(-135,120){\large \underline{Type $3$}}
        }
   \caption{The three configurations types for the deployment of component \ModB{u} of a fog application.}\label{fig:conf}
\end{figure*}
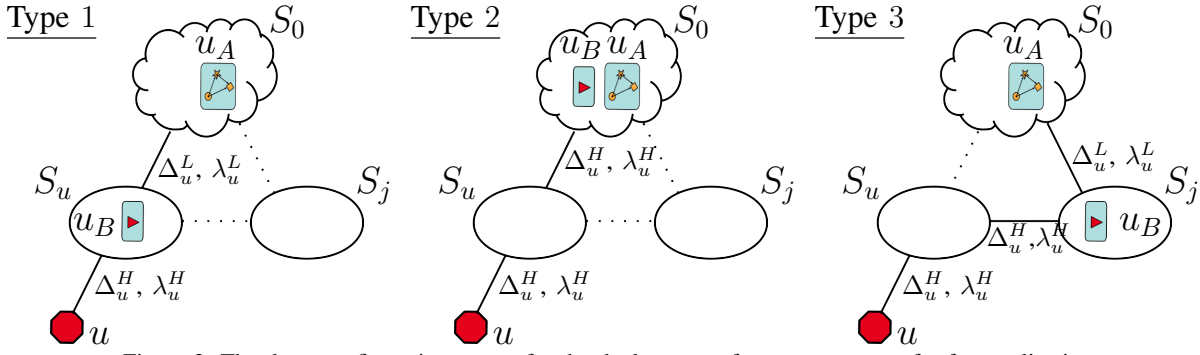

{\noindent \em Application Architecture.} As depicted in Fig.~\ref{fig:conf}, an application $u \in \U$ consists of two
containers: \ModA{u} and \ModB{u}. In order to account for computing and communication constraints in a practical case,
we refer to a benchmark application for face recognition in a video stream. As introduced before, modules for processing
IoT data streams -- face detection processing over the sequence of video frames in our example  -- are containerized
in \ModB{u}. They can be deployed in the central cloud $\Servers{0}$ or on the edge, i.e., in regions $\Servers{i}, i=1,2,3$.
Conversely, \ModA{u} contains all remaining logic, including, e.g., alarm generation in case a positive match is returned.
The application has to output every $1/F_u$ seconds a result $y_u$ -- in this case a positive or negative face recognition
match. \ModA{u} is installed in the central cloud $\Servers{0}$. We can hence consider the whole processing chain involved
by the two-containers and the related data transmission delay. We should also include the processing delay $d_u$ of
application $u$ (if deployed back to back to the IoT object), plus the communication delay $d_{uj}$, which is the additional
delay to retrieve data from region where the sensor belongs to \ModA{u}, when \ModB{u} is installed in region $j$.

The IoT source -- in the example a videocamera -- generates information units -- video frames -- of size $\Delta_u$, which
are served at rate $B_u$ bit/s. We denote $\DataH{u}=\Delta_u$. Conversely, \ModB{u} transfers smaller information unit
\DataL{u} to \ModA{u}. 

Finally, we denote \MemApp{u}, \StorApp{u}, \ProcApp{u} the resource requirements of of application $u$, in
terms of memory, storage and processing capacity, respectively, of \ModB{u}; with compact notation we denote
$\ReqApp{u}=(\MemApp{u}, \StorApp{u}, \ProcApp{u})$

In the placement problem we need to consider the processing and transferring time. Actually, the processing time for
each information unit depends on the throughput between application modules. Any application placement has to
guarantee that the application to process an information unit \Data{u} in $\frac{1}{\Freq{u}}$ seconds. Thus, the
allocation of such throughput depends on the app deployment configurations. Since \ModA{u} is always installed on
the central cloud, the three basic fog configurations to deploy app $u$ are as in Fig.~\ref{fig:conf}:\\
{\noindent \em Type 1:} \ModB{u} deployed on $\Servers{u}$; higher throughput \ThroughH{u} flows between IoT object $u$ and region $\Servers{u}$, with IoT data unit $\Data{u} = \DataH{u}$. \DataL{u} is served between \Servers{u} and \Servers{0} with low throughput \ThroughL{u};\\
{\noindent \em Type 2:} \ModB{u} deployed on central cloud $\Servers{0}$;  the IoT data $\Data{u} = \DataH{u}$ is served between \Servers{u}and \Servers{0} with high throughput \ThroughH{u};\\
{\noindent \em Type 3:} \ModB{u} deployed on a neighboring fog region $\Servers{j}\not=\Servers{u}$;
lower throughput required between \Servers{j} and central cloud \Servers{0}. However, the IoT data $\Data{u} = \DataH{u}$
is served between \Servers{u} and \Servers{0} with high throughput \ThroughH{u}.


\section{Problem formulation}\label{sec:probform}


The resource allocation problem is tackled from the perspective of the edge-infrastructure owner. Her aim is to
maximize the revenue obtained in the provision of her fog infrastructure to application tenants. In fact, she settles
a cost in order to deploy an application using the traditional scheme of pay per use. A tenant owning application
$u$ pays $f_{u,k}>0$ euros per container installed in region $k$.

The objective is to schedule the containerized fog applications such in a way to maximize the owner revenue,
while satisfying the applications' requirements. We can obtain the optimal reward for a given set of application
requests. Hence, the following formulation provides an upper bound on the average reward that can be attained with
perfect information.

Decision variables $x_{u,k,i}$ are boolean variables indicating the placement of the application $u$ on the $i$-th server
of the region $k$. Further, decision variables $\ThroughH{u}, \ThroughL{u} \in \mathbb{R}^+$ represent throughput in the
large and small data unit transfer mode of application $u$, respectively. The optimal allocation policy using
a mixed integer non linear program (MINLP) writes:

\begin{small}
\begin{align}
  \mbox{maximize:}&\quad \sum_{(u,k) \in \U \times \K \setminus \{u\}} f_{u,k} \, x_{u,k}  \label{prob1}\\
  \mbox{subject to:} \nonumber \\
&\sum_{u \in \U} \ReqApp{u} \,x_{u,k,i} \le \UnitV{k}{i}, \quad  \forall k \in \K, \forall i \in \Servers{k} \label{eq:reqApp}\\   
&\sum_{u \in \Apps{k}} (x_{u,k} \,\ThroughL{u} + x_{u,0} \,\ThroughH{u}) + \nonumber\\
&+\sum_{j \in \N(\Servers{k})} \sum_{v \in \Apps{j}} x_{v,j} \ThroughL{v} \le \BandR{k}{0}, \quad \forall k \in \K \setminus \{0\} \label{eq:bwToCloud}\\
&\sum_{u \in \Apps{k}} x_{u,j} \,\ThroughH{u} + \sum_{u \in \Apps{j}} x_{u,k} \,\ThroughH{u} \le \BandR{k}{j} ,\; \forall jk \in E, j,k \ne 0 \label{eq:bw3_1}\\
&\DelayA{u} + \frac{\DataH{u}}{\BandA{u}} + \Big ( \DelayR{u}{j} + \frac{\DataH{u}}{\ThroughH{u}} + \frac{\DataL{u}}{\ThroughL{u}} \Big) \,x_{u,j} + \nonumber\\
&\Big ( \DelayR{u}{0} + \frac{\DataH{u}}{\ThroughH{u}} \Big) \,x_{u,0} + \Big ( \DelayR{u}{0} + \frac{\DataL{u}}{\ThroughL{u}} \Big) \,x_{u,u} \le \frac{1}{\Freq{u}} \nonumber\\
&\forall u \in U, \forall j \in \N(\Servers{u})\label{eq:transm}\\
&\sum_{k \in \K} x_{u,k} \le 1 \quad \forall u \in \U \label{eq:unique}\\
&\sum_{k \in \K \setminus \{\N(u) \cup \{u\}\}} x_{u,k} \le 0 \quad \forall u \in \U \label{eq:onehop}\\ 
&x_{u,k,i} \in \{0,1\} \quad \forall (u,k) \in \U \times \K \quad \forall i \in \Servers{k}\\
&\ThroughH{u}, \ThroughL{u} \in \mathbb{R}^+ \label{final}
\end{align}
\end{small}
where we let $x_{u,k} = \sum_{i \in \Servers{k}} x_{u,k,i} \quad \forall (u,k) \in \U \times \K$ for notation's sake. 
The objective function is the revenue gained by the infrastructure owner. 
The constraint~\eqref{eq:reqApp} is meant component-wise: it bounds the resources utilization on fog servers in terms of memory, processing and storage capacity, respectively. Also,~\eqref{eq:bwToCloud} and~\eqref{eq:bw3_1} bound the
throughput generated by applications with respect to links' capacity. ~\eqref{eq:bwToCloud} accounts for all traffic from
region $k$ to the central cloud, whereas~\eqref{eq:bw3_1} accounts for the throughput across adjacent regions as in ~\ref{fig:conf}c. By constraint~\eqref{eq:transm}, the total transmission and computing time needs to be smaller than the service rate of the application. We assume that, according to~\eqref{eq:unique}, each application has at most one deployment region. In particular, ~\eqref{eq:onehop} indicates that each application can be deployed only on neighbor regions or on its original region.

The decision variables are the binary variables for the placement and the continuous variables for the throughput. The
 Prob.~\ref{prob1}--\ref{final} is a combination of a placement problem and a multicomodity flow problem. For the sake of tractability, in the next section we offer a reduction to a pure placement problem, which is seen to correspond to a $m$-dimensional knapsack problem.

\section{Pure placement problem}\label{sec:place}

The reduction is attained by fixing the continuous decision variables of the MINLP, i.e., $\ThroughL{u}$ and $\ThroughH{u}$.
To do so, we fix the minimum throughput required for each application $u \in \U$ to deliver the output at target rate
$\Freq{u}$, given the configuration type and the deployment region for \ModB{u}.

{\noindent \em Type. 1:} processing each information unit and providing an output result should happen
at rate $\frac{1}{\Freq{u}}$; by accounting for all processing and communication delay we write
\begin{equation} \label{eq:1}
\DelayA{u} + \DelayR{u}{0} + \frac{\DataH{u}}{\BandA{u}} + \frac{\DataL{u}}{\ThroughL{u}} \le \frac{1}{\Freq{u}}
\end{equation}
which can be solved for equality in \ThroughL{u};\\
{\noindent \em Type. 2:}  For each application $u$, we have
\begin{equation}
\DelayA{u} + \DelayR{u}{0} + \frac{\DataH{u}}{\BandA{u}} + \frac{\DataH{u}}{\ThroughH{u}} \le \frac{1}{\Freq{u}}
\end{equation}
In this case we are solving for \ThroughH{u}; we observe that it must hold indeed $\ThroughH{u} \geq \ThroughL{u}$.\\
{\noindent \em Type. 3:} if \ModB{u} is deployed in a region neighbor of the original region of $u$, it holds 
\begin{equation}\label{eq:9}
\DelayA{u} + \DelayR{u}{j} + \DelayR{j}{0} + \frac{\DataH{u}}{\BandA{u}} + \frac{\DataH{u}}{\ThroughH{u}} + \frac{\DataL{u}}{\ThroughL{u}} \le \frac{1}{\Freq{u}}
\end{equation}
In this case, in order to have a unique solution in the minimum throughout, we impose additional constraints, namely we restrict to the set of solutions such that 
\begin{equation}
\frac{\ThroughH{u}}{\ThroughL{u}} = \frac{\DataH{u}}{\DataL{u}}
\end{equation}

Once we performed the above identification, the original problem becomes:
\begin{small}
\begin{align}
\mbox{maximize:}  & \sum_{(u,k) \in \U \times \K} f_{u,k} \,x_{u,k} \label{eq:objk}\\
\mbox{subject to:} \nonumber \\
&\sum_{u \in \U} \ReqApp{u} \,x_{u,k,i} \le \UnitV{k}{i}, \quad  \forall k \in \K, \forall i \in \Servers{k} \label{eq:reqAppk}\\
&\sum_{u \in \Apps{k}} (x_{u,k} \,\ThroughL{u} + x_{u,0} \,\ThroughH{u}) + \nonumber\\
&+\sum_{j \in \N(\Servers{k})} \sum_{v \in \Apps{j}} x_{v,j} \ThroughL{v} \le \BandR{k}{0}, \quad \forall k \in \K \setminus \{0\} \label{eq:bwToCloudk}\\
&\sum_{u \in \Apps{k}} x_{u,j} \,\ThroughH{u} + \sum_{u \in \Apps{j}} x_{u,k} \,\ThroughH{u} \le \BandR{k}{j} , \; \forall jk \in E, j,k \ne 0 \label{eq:bw3_1k}\\
&\sum_{k \in \K} x_{u,k} \le 1 \quad \forall u \in \U \label{eq:uniquek}\\
&\sum_{k \in \K \setminus (\N(u) \cup \{u\})} x_{u,k} \le 0 \quad \forall u \in \U \label{eq:onehopk}\\
& x_{u,k,i} \in \{0,1\}, \quad \forall (u,k) \in \U \times \K, \quad \forall i \in \Servers{k} \label{eq:final} 
\end{align}
\end{small}
\begin{prop}
Problem \eqref{eq:objk} is \textit{NP}-hard.
\end{prop}
\begin{IEEEproof}
For every instance of a multidimensional knapsack with $n$ decision variables and $m$ constraints, we can reduce it to an instance of our problem. In fact, it is sufficient to consider an instance of \eqref{eq:objk}--\eqref{eq:final} with $n$ applications and a single $m$ servers region, which proves $NP$-hardness. 
\end{IEEEproof}
We note that \eqref{eq:objk}--\eqref{eq:final} appears as a $m$-knapsack instance, where $m = K \sum_{k \in \K} n_k + |E| + 2U$: in the decision form, the problem is hence $NP$-complete.

\subsection{Placement algorithm}\label{sec:greedy}


Hereafter, we describe FPA, a greedy solution for \eqref{eq:objk}. 
\begin{algorithm}
\begin{small}
\Input{$G = (V,E)$, $\U$}
\Output{Container placement for each $u \in \U$}
\While{$\U \ne \emptyset$}{
	\For{$i = 1,\dotsc,K$}{
		\For{$u \in \Apps{i}$}{
			$\mathcal{A} \leftarrow \emptyset$\;
			\If{$verify(\Servers{i}, u) = TRUE$}{
				$\mathcal{A} \leftarrow \mathcal{A} \cup \{\Servers{i}\}$\;
			}
			\For{$S \in \N(S)$}{
				\If{$verify(S, u) = TRUE$}{
					$\mathcal{A} \leftarrow \mathcal{A} \cup \{S\}$\;
				}
			}
			\If{$|\mathcal{A}| \ge 2$}{
				$(j^*, s^*_{j_{h^*}}) \leftarrow select(\mathcal{A}, $u$)$\;
				\tcp{where $s^*_{j_{h^*}} \in \Servers{j^*}$}
			}
			\ElseIf{$|\mathcal{A}| = 1$}{
				$(j^*, s^*_{j_{h^*}}) \leftarrow \Servers{j^*}$ with $\Servers{j^*} \in \mathcal{A}$\;
			}
		}
	}
	\tcp{select the application to be deployed}
	$u^* \leftarrow \argmin\limits_{u \in \U} {\left\lVert \bar{v}_{j^*}^u \right\rVert}^2$\;
	deploy($u^*, j^*$)\;
	updateServer($\Servers{j^*}, s^*_{j_{h^*}}, u^*$)\;
	\tcp{Update $G$}
	update($G, \Servers{j^*}, \Servers{u^*}, u^*$)\;
	$\U \leftarrow \U \setminus \{u^*\}$
}
\caption{Fog Placement Algorithm (FPA)}
\label{algo:alg1}
\end{small}
\end{algorithm}

FPA operates an iterative application deployment. At each step, for each region and for each application $u$ which belongs to that region, it selects the set $\mathcal{A}$ of admissible regions for the deployment of module \ModB{u} container. Such set includes all the regions satisfying the computational and throughput requirements of a tagged application. Preliminarily, a feasibility check is performed through a \textit{verify} procedure (pseudocode omitted for space's sake): given a region and application's requirement, it verifies whether exists some server in the region to host $\ModB{u}$. Further, throughput requirements are verified against each configuration type for each application, by ensuring that the residual bandwidth of involved links satisfies the minimum throughput requirement corresponding to the tagged configuration type.

The \textit{select} procedure is reported in Algo.~\ref{algo:select}: \textit{select} first calculates, for all eligible applications to be still deployed, a gradient $\bar{v}_S$ for each feasible region. Its components are calculated at lines 1, 2, 3, 7-8, 11, and 14, respectively, by estimating  the normalized decrease of each resource type in case of deployment with tagged configuration. The output is the application minimizing the gradient (line 16).
\begin{algorithm}
\begin{small}
\Input{$\mathcal{A}$, set of admissible regions for the deployment of the module $\ModB{u}$}
\Output{A region for the deployment}
\tcp{Build a gradient vector for each region in $\mathcal{A}$}
\For{$S \in \mathcal{A}$}{
	$v_m \leftarrow \frac{\MemApp{u}}{residual\_mem(S)}$\;
	$v_p \leftarrow \frac{\ProcApp{u}}{residual\_proc(S)}$\;
	$v_s \leftarrow \frac{\StorApp{u}}{residual\_stor(S)}$\;
	\eIf{$S \ne \Servers{u}$}{
		\eIf{$S \in \N(\Servers{u})$}{
			\tcp{Case 3}
			$b_1 \leftarrow \frac{\ThroughH{u}}{residual\_band(\{u,S\})}$\;
			$b_2 \leftarrow \frac{\ThroughL{u}}{residual\_band(\{S,0\})}$\;
			$\bar{v}_S \leftarrow (v_m, v_p, v_s, b_1, b_2)$\;
		}{
			\tcp{$S = \Servers{0}$, case 2}
			$b_1 \leftarrow \frac{\ThroughH{u}}{residual\_band(\{0,u\})}$\;
			$\bar{v}_S \leftarrow (v_m, v_p, v_s, b_1, 0)$\;

		}
	}{
		\tcp{Case 1}
		$b_1 \leftarrow \frac{\ThroughL{u}}{residual\_band(\{0,u\})}$\;
		$\bar{v}_S \leftarrow (v_m, v_p, v_s, b_1, 0)$\;
	}
}
\Return{$\argmin \limits_{S \in \mathcal{A}} \{ {\left\lVert \bar{v}_S \right\rVert}^2 \}$}
\caption{\textit{Select} procedure}
\label{algo:select}
\end{small}
\end{algorithm}
Once the algorithm has selected the application to be deployed, it updates the computational capacities of the server hosting the module of that application.
Afterwards, the algorithm updates the graph structure decreasing the bandwidth of the links that connected the regions selected for the deployment (line 17). It iterates until all applications have been considered.

{\noindent \em Complexity.} Now we look at the complexity of FPA. The procedures \textit{verify}, \textit{updateServer} and \textit{update} have constant time complexity. The procedure \textit{select} computes a vector for each eligible region in the set $\mathcal{A}$. In the worst case, the cardinality of $\mathcal{A}$ is at most $K-1$. Hence, the complexity of the \textit{select} procedure is $O(K)$. The cardinality of $\U$ is $O(U)$, and the maximum cardinality of a neighborhood of a certain region is $O(K)$ in the worst case. Finally, the complexity of FPA is $O(U^2 \cdot K^3)$.


\section{Real Implementation: FogAtlas}\label{sec:fogatlas}

FPA is the fog scheduler of FogAtlas, a fog platform derived from several extensions of
the early platform described in~\cite{Foggy2017}. It handles microservice deployment and workload
placement by managing a distributed fog infrastructure split into one cloud region and one or more fog 
regions. Actually, FogAtlas has a region-oriented architecture. In fact, existing OpenSource
technologies such as OpenStack and Kubernetes handle well resources orchestration in traditional data centers
where the cloud is centralized (optionally also spread across few large regions). However, they do not
handle natively distributed and/or decentralized fog systems, where heterogeneous computing devices
lay in diverse IoT regions and must be internetworked with a central cloud, often with bandwidth-limited
and/or partially reliable connections. Ultimately, FogAtlas handles the orchestration among regions, while
delegating intra-region orchestration to standard OpenStack or Kubernetes controllers. 

\begin{figure}[t]
        \centering
        \includegraphics[width=\linewidth]{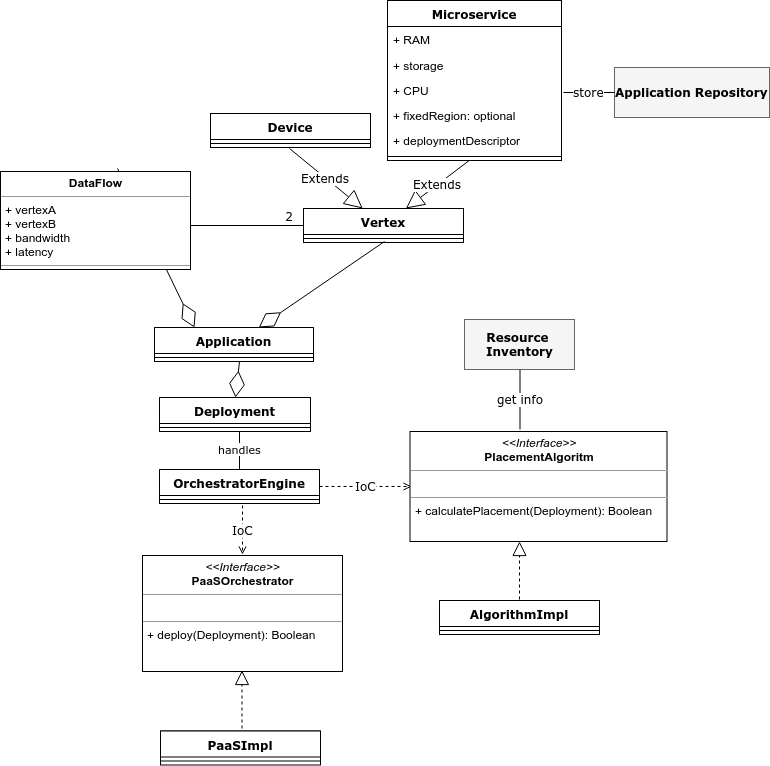}
   \caption{The FogAtlas \texttt{Orchestrator} and its implementation}\label{fig:fogatlas}
\end{figure}

The platform instantiates fog applications accouting for a set of optional
deployment \emph{requirements}. The application owner can specify requirements
as constraints imposed to the deployment/execution of microservices in terms of 
requested resources and/or specific application needs. She is allowed to declare
connections of IoT objects with a certain \texttt{Microservice}, see Fig~\ref{fig:fogatlas}.
She can also require a specific target region for dispatching. 

In this context \texttt{Microservice} is a unit of software which plays
a specific role as part of a larger fog application. But it can be deployed,
upgraded or replaced independently from other microservices of same application.
In FogAtlas it is distributed via Docker container images, which are stored in an
\texttt{Application Repository}, in fact a Docker registry.

FogAtlas adds above OpenStack and Kubernetes an \texttt{Orchestrator}, an \texttt{Inventory},
a \texttt{Monitor} and a set of RESTFul API together with some other components needed to
operate the whole platform.

{\noindent \bf FogAtlas Inventory.} The~\texttt{Inventory}
maintains an annotated topology of the distributed infrastructure and the applications deployed
with up-to-date information on the state of resources.  The \texttt{Inventory} maps infrastructural
objects (i.e., regions, nodes, things) and application objects (i.e. applications, microservices)
keeping track of their location and deployment status. As far as the infrastructural objects are
concerned, the \texttt{Inventory} is populated with information from external systems like SDN
network orchestrators and/or IaaS managers (e.g., ONOS, OpenStack). On the other hand, application
related information is taken from PaaS managers (i.e., Kubernetes). Information is maintained based
on a distributed and highly available key value store~\cite{etcd}.

{\noindent \bf FogAtlas Orchestrator.} The \texttt{Orchestrator} (see Fig~\ref{fig:fogatlas}a) receives \texttt{Deployment} requests
referred to an \texttt{Application} and try to place related \texttt{Microservice} in a way that best satisfies the imposed requirements.
An  \texttt{Application} is modeled as a graph of \texttt{Vertices} (\texttt{Devices} or \texttt{Things} used by the \texttt{Application}
and \texttt{Microservices}) and \texttt{Dataflows}. Both  \texttt{Vertices} and \texttt{Dataflows} can specify requirements in terms of usage
of resources and geographical location. We use \emph{Inversion of Control} design principle in order to inject into the \texttt{Orchestrator}
the specific implementation of the placement algorithm and of the PaaS manager in use (i.e. Kubernetes).

We remark that in FogAtlas we support geographical constraints (regions) and bandwidth
constraints which are not standard features of traditional cloud schedulers. The
\texttt{Application Repository} is a Docker registry, typically deployed on the cloud
tier, and contains the application images, i.e., \texttt{Microservice} components. 

The application deployment is performed as follows. A deployment request
is submitted using the FogAtlas RESTful API. Requests can be processed in batches
 or sequentially (unitary batch). The first step is performed by the \texttt{Orchestrator}: it queries the 
\texttt{Inventory}, applies filtering and ranking rules as defined by the \texttt{PlacementAlgorithm} to identify the best
regions to host the \texttt{Microservice} of the requested \texttt{Application}. Regions satisfying the requirements
specified in the deployment request are identified: hence, the \texttt{Orchestrator} operates according to the results
of the \texttt{PlacementAlgorithm}. The \texttt{PaasOrchetstrator} finally deploys on the target region the container
image of the \texttt{Microservice}. The actual deployment of the \texttt{Microservice} on a node of the selected region
is left to the PaaS manager (in this case Kubernetes). At the end of the process, the FogAtlas monitor component updates
the \texttt{Inventory} to reflect the global status of infratructure resources.

\subsection*{FogAtlas Implementation} We provide hereafter a few technological
details on FogAtlas. In order to combine IaaS availability with flexible management
of edge nodes, in FogAtlas the IaaS layer is provided by OpenStack while
Kubernetes performs container orchestration. In particular, the OpenStack
deployment adheres to the architecture proposed by th Edge Computing Group
\cite{femdc}. Namely, the OpenStack controller lies in the cloud tier while compute nodes cover the edge
devices. They are interconnected via "WANWide" links. A Kubernetes cluster
is distributed on top of OpenStack virtual machines, covering both cloud and
edge nodes. In case of small edge devices (with respect to available resources)
OpenStack is not installed and Kubernetes workers are deployed directly on bare metal.

\noindent \emph{Physical testbed and measurements:} the FBK data center
holds the cloud tier and the edge cloudlet tier. Server nodes mount an Intel i7 CPU,
16GB RAM, and 480GB SSD. Furthermore, dedicated edge gateways can connect small and
low power consumption devices (Raspberry Pi version 3), to perform hardware abstraction
layer and to connect for non-IP IoT devices. TP-Link TL-WR740N access points and Tenvis
JPT3815W-HD cameras are finally connected to our plate recognition application \cite{Foggy2017}.

In order to provide realistic scenarios for our numerical evaluation, we have measured 
resources demands of such benchmark application (see Tab.~\ref{tab:appReq}). In the same way,
placement constraints due to server characteristics (memory, CPU and storage) do mimic current
expected consumer electronics specifications, FogAtlas servers (see Tab.~\ref{tab:servers}). The
objective is to test the scalability of our fog placement mechanism with the applications batch size,
as described in the next section.


\section{Numerical Results}\label{sec:numres}


First, we describe the setup of the tested scenarios. Where not otherwise specified, we intend the infrastructure
owner to maximize the number of deployed applications, i.e., $f_{u,k}\equiv 1$.

\textit{Network topology:} we consider a reference undirected network graph with a fixed number of regions $K=10$,
where the central cloud and regions form a star topology of cloud-to-fog connections, namely cloud-links. For every topology realization,
crosslinks among regions are added according to an Erd\"os Renyii random graph model, where a link exists between
two regions with probability $q$. Finally, we assign to each link in the resulting network a bandwidth of $15$ Mbps,
both for the cloud-links and crosslinks.

\textit{Application Batch Generation:} a batch of fog applications is generated for each experiment; we considered 
$U=\{10, 50, 100, 150, 250\}$. The demands of each application of the batch for CPU, storage, memory and throughput
are uniform independent random variables. The mean value of such variables is dictated by the nominal value we
measured for our benchmark application. That application, as recalled in the previous section, is a plate-recognition
application packaged as a two-modules microservice. The second microservice module can process the video
stream either in the cloud or on a fog node. The resulting distribution values for the application batches are
enlisted in Tab.~\ref{tab:appReq}; symbol $u_0$ refers to the nominal values we measured on
FogAtlas for the plate recognition app.

Finally, the probability that an application belongs to region $k \in  \{1,\ldots,K\}$ follows a
truncated Pareto distribution of parameter $\alpha$, i.e., $\mathbb{P}\{R_u > k\} = {k}^{-\alpha}/\gamma$, 
where $R_u$ is the random variable representing the index of the region assigned to the application $u$
and normalization constant $\gamma=\sum_{h=1}^K {h}^{-\alpha}$. 

\textit{Fog Server Classes:} the servers available within each region belong to three classes, depending on
the resources they are equipped with, namely \textit{low}, \textit{medium} and \textit{high} class. The computational
characteristics are listed in Table~\ref{tab:servers}. The number of servers per region is determined per realization
as follows. Each region is meant to satisfy same fraction of the expected aggregated demand. More precisely, each region
is equipped with aggregated resource  vector $(1+\beta) \frac{U}{K} \ReqApp{u_0}$. The parameter $\beta$ is a slack
parameter tuning the probability that fog resources are underprovisioned/overprovisioned compared to the aggregated demand.
Finally, the servers' population of the tagged region is determined by allocating servers of
random type until the region resource budget is exhausted.

\begin{table}[t]\caption{Distribution of the application requirements of CPU, memory, storage and throughput.}
\centering
\begin{tabular}{|p{0.30\columnwidth}|p{0.25\columnwidth}|p{0.30\columnwidth}|}
\hline
{\it Requirement} & {\it Mean Value ($u_0$)} & {\it Range ($u \in \U$)} \\
\hline
CPU (\ProcApp{u}) & 1250 MIPS & $[500, 2000]$  MIPS \\
\hline
Memory (\MemApp{u})& 1.2 Gbytes & $[0.5, 2]$ Gbytes \\
\hline
Storage (\StorApp{u})& 3.5 Gbytes & $[1, 8]$ Gbytes \\
\hline
Low throughput (\DataL{u}) & $1.5$ Mbps &  $[1, 2]$ Mbps \\
\hline
High throughput (\DataH{u}) & 4.25 Mbps &  $[3.5,5]$ Mbps \\
\hline
\end{tabular}\label{tab:appReq}
\end{table}
\begin{table}[t]\caption{Characteristics of the three classes of fog servers: low, medium and high.}
\centering
\begin{tabular}{|p{0.1\columnwidth}|p{0.2\columnwidth}|p{0.2\columnwidth}|p{0.2\columnwidth}|}
\hline
{\it Type} & {\it CPU (MIPS)} & {\it Memory (GB)} & {\it Storage (GB)}\\
\hline
Low &  5000 & 2 & 60\\
\hline
Medium & 15000 & 8 & 80\\
\hline
High & 44000 & 16 & 120\\
\hline
\end{tabular}\label{tab:servers}
\end{table}

\subsection{Experimental Results}

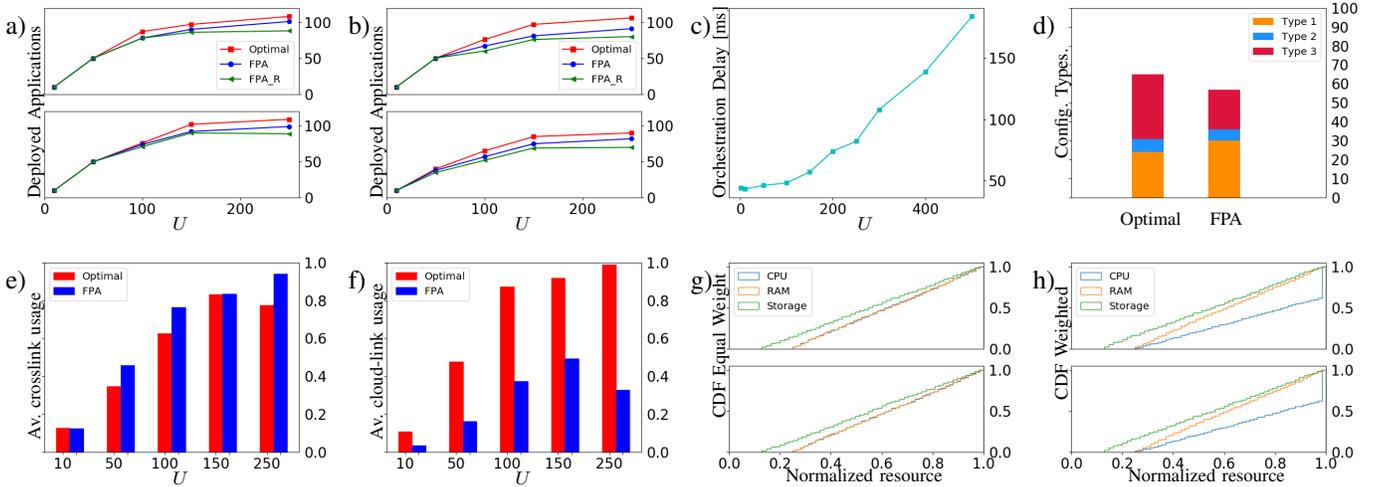
\begin{figure*}[t]
  \begin{center}
    \includegraphics[width=0.24\linewidth]{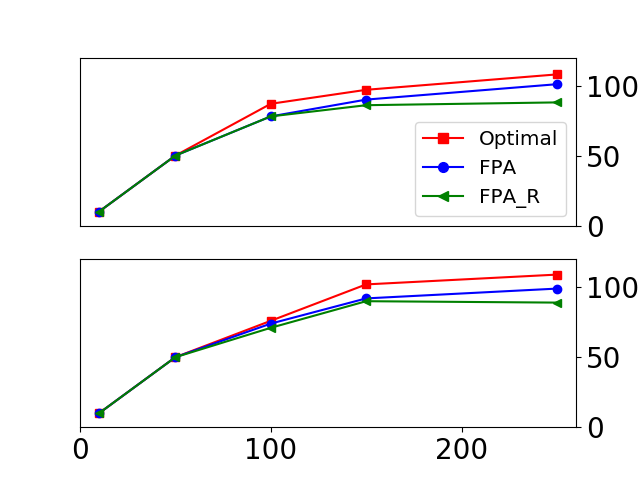}
    \put(-60,-2){\scriptsize $U$}\put(-110,10){\begin{rotate}{90}{\scriptsize Deployed Applications}\end{rotate}}\put(-123,72){a)}\hskip2mm
    \includegraphics[width=0.24\linewidth]{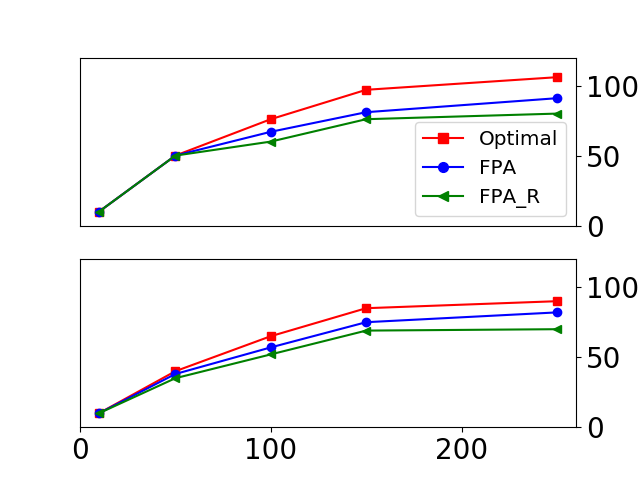}
    \put(-60,-2){\scriptsize $U$}\put(-110,10){\begin{rotate}{90}{\scriptsize Deployed Applications}\end{rotate}}\put(-123,72){b)}\hskip2mm
    \includegraphics[width=0.24\linewidth]{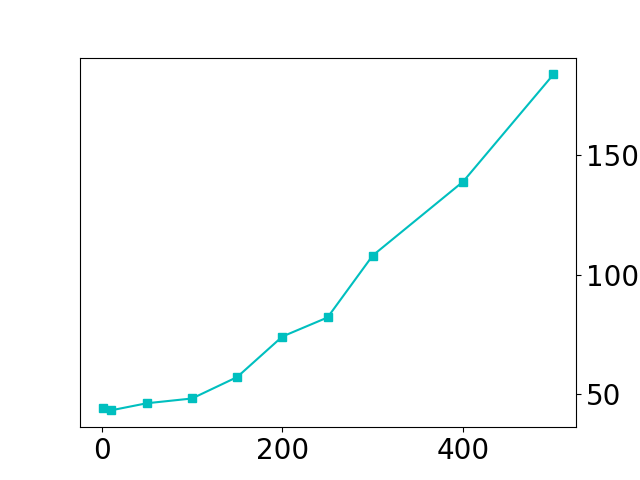}
    \put(-60,-2){\scriptsize $U$}\put(-110,10){\begin{rotate}{90}{\scriptsize Orchestration Delay [ms]}\end{rotate}}\put(-123,72){c)}\hskip2mm
    \includegraphics[width=0.24\linewidth]{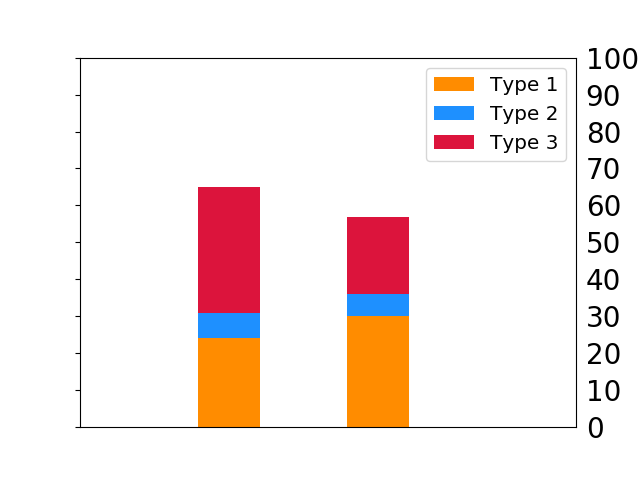}
    \put(-56,0){\scriptsize FPA}\put(-90,0){\scriptsize Optimal}\put(-110,25){\begin{rotate}{90}{\scriptsize Config. Types.}\end{rotate}}\put(-123,72){d)}\\[0.5mm]
    \includegraphics[width=0.24\linewidth]{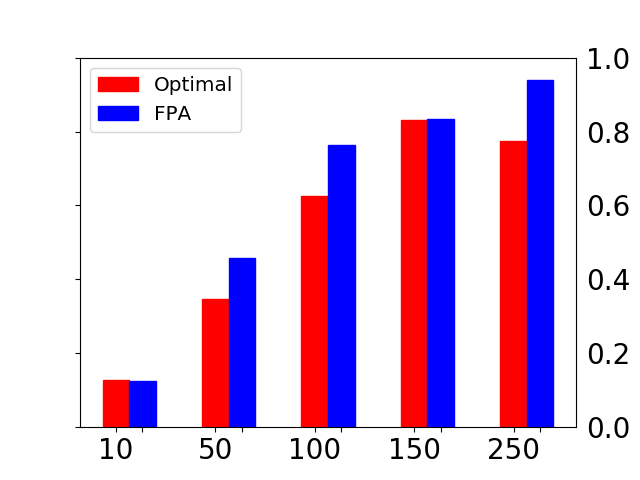}
    \put(-60,-2){\scriptsize $U$}\put(-110,15){\begin{rotate}{90}{\scriptsize Av. crosslink usage}\end{rotate}}\put(-123,72){e)}\hskip2mm
    \includegraphics[width=0.24\linewidth]{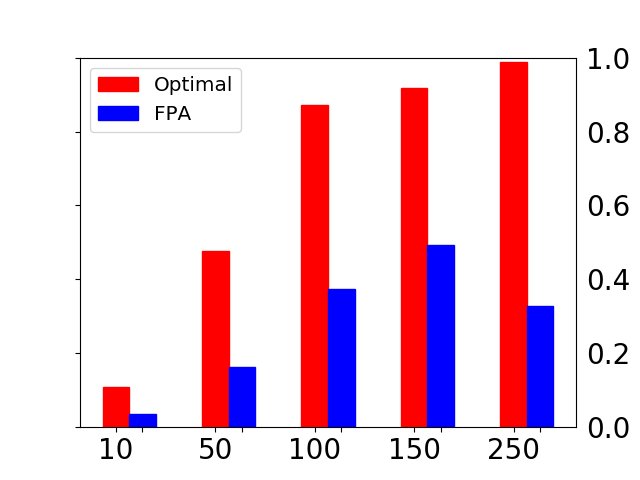}
    \put(-60,-2){\scriptsize $U$}\put(-110,15){\begin{rotate}{90}{\scriptsize Av. cloud-link usage}\end{rotate}}\put(-123,72){f)}\hskip2mm
   \includegraphics[width=0.24\linewidth]{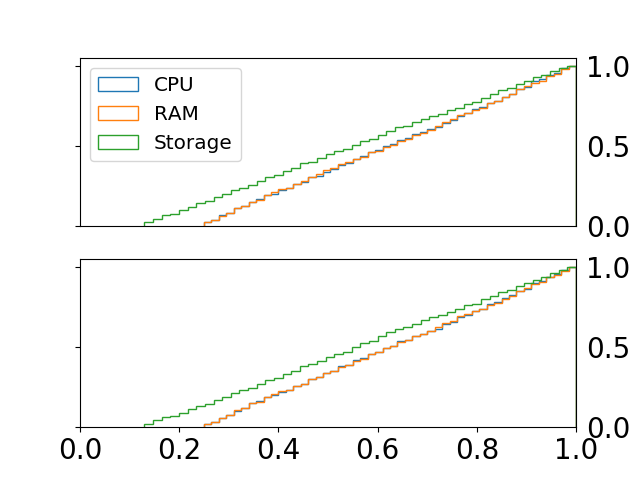}
    \put(-87,-1){\scriptsize Normalized resource}\put(-110,23){\begin{rotate}{90}{\scriptsize CDF Equal Weight}\end{rotate}}\put(-123,72){g)}\hskip2mm
    \includegraphics[width=0.24\linewidth]{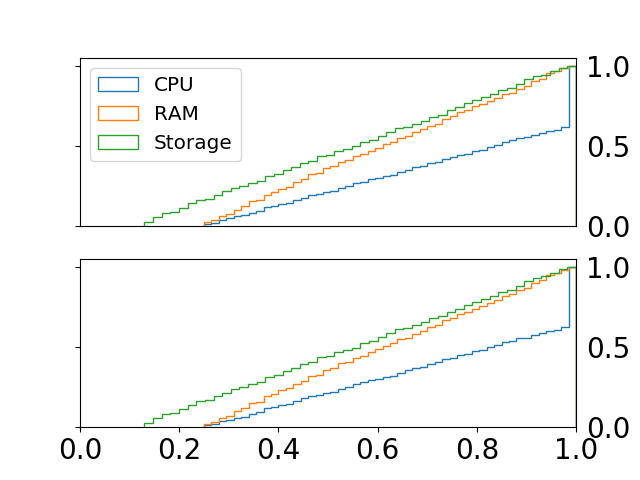}
    \put(-87,-1){\scriptsize Normalized resource}\put(-110,30){\begin{rotate}{90}{\scriptsize CDF Weighted}\end{rotate}}\put(-123,72){h)}\hskip4mm
  \end{center}
  \caption{a/b) Number of deployed applications: a) $q=0.4$ $\beta=1.5$ (top) and $\beta=2.5$ and (bottom); %
    b) $\beta=0.3$ and $q=0.5$ (top) and $q=0.3$ (bottom); %
    c) Orchestration delay, $\beta=0.3$ and $q=0.5$(top) and $q=0.3$; %
    d) Configuration types distribution for a typical solution instance with $U=100$, $q=0.3$ and $\beta=0.3$; %
    e/f) Average link usage (settings as in d): e) cloud-links and  %
    f) crosslinks; %
    g/h)   CDF of the demands for the deployed applications g) Equal weight, optimal and FPA solutions, $q=0.5$ and $\beta=1.5$ %
       and h) Weighted, optimal and FPA solutions, $q=0.5$ and $\beta=0.5$;
  }\label{fig3}    
\end{figure*}

In Fig.~\ref{fig3}a we have depicted the number of deployed applications for increasing batch size. The upper
graph reports the results averaged on $10$ instances of a scenario with parameter $\beta=1.5$ (top) and $\beta=2.5$
(bottom), respectively. The red line is the optimal solution obtained by the Gurobi ILP solver~\cite{gurobi},
the blue line is FPA, whereas the green one is the variant of FPA implemented in FogAtlas, namely FPA-R. It 
considers region-wise aggregated resources and delegates the intra-region, per-server deployment to Kubernetes
schedulers using a randomized placement policy%
\footnote{Basically, the algorithm runs FPA as if there exists a unique server having aggregated capacity of the entire region.}. As seen in Fig.~\ref{fig3}a, up to $U=50$, the deployment of the batch of applications is complete. In the last part of the curve, communication constraints dominate, saturating around $100$ deployed applications in the optimal case. Increasing from $\beta=1.5$ to $\beta=2.5$ provides moderate improvement, confined around $U=100$, where the communication constraint is not dominating yet. 

Fig.~\ref{fig3}b repeats the same experiment in the case of different crosslink density among regions. The figure on top
represents the case of denser topologies ($q=0.5$) and the bottom one the case of sparser ones ($q=0.3$). We observe first that using $\beta=0.3$, and $q=0.5$ (top graph), this scenario has close performance to the ones seen in Fig.~\ref{fig3}a, but for much lesser computational resources assigned to fog regions. However, when the network is sparser (bottom graph), the demand peaks for regions of lower indexes -- according to the Pareto distribution -- are not offloaded to neighboring regions. This causes the bottleneck visible even for smaller batch sizes, i.e., $U=10,50$.

From Fig.~\ref{fig3}a and b we observe that for the chosen settings, FPA has performance close to the optimal
solution, whereas FPA-R pays some performance loss which is traded off for implementation's simplicity.

Fig.~\ref{fig3}c reports on the tests performed on the orchestration delay on the FogAtlas platform, defined as the time
needed from the instant when the batch of application is offered to the scheduler until the placement is calculated. As
we can see, the expected time complexity is moderately super-linear, confirming scalability to larger batch sizes.

We tested again the sparser deployment ($q=0.3$) already described in Fig.~\ref{fig3}b, for $U=100$. In
Fig.~\ref{fig3}d, we have generated a typical instance and described the configurations of the deployments produced
by FPA and by the optimal solution. The latter prefers type 3 configurations over type 1 configurations, whereas the opposite occurs for FPA. The impact onto the link utilization is different: we tested the link utilization in Fig.~\ref{fig3}e and f. Actually, crosslinks are fully utilized in both cases, see Fig.~\ref{fig3}e. But, offloading using Type 3 configurations is less frequent with the greedy algorithm: in turn cloud-links are underutilized (Fig.~\ref{fig3}f). The different behaviour is due to the fact that, in a throughput-dominated scenario, optimal solutions prioritize communication constraints more efficiently than FPA's ones. 

Finally, Fig.~\ref{fig3}g and h characterize deployed applications for different weights. We depicted there the Cumulative Distribution Function (CDF) for the memory, storage and CPU required by the selected applications. The distribution is uniform in the case of equal weights, indicating that both optimal and FPA solutions sample applications to deploy uniformly at random with respect to computing requirements. This is what desired in a throughput-dominated scenario, proving that FPA behaves correctly by prioritizing the communication constraints. In the second scenario, half applications are generated with the maximum CPU value and the others uniform. We have assigned to each application $u$ the weight $\frac{\ProcApp{u}}{max\_CPU}$, i.e., according to their probability mass distribution. Doing so, both the optimal and the FPA solutions have deployed applications according to the weight distribution, prioritizing higher CPU consumption.


\section{Related Work}\label{sec:related}


Efficient service deployment is a core topic in cloud computing~\cite{cohen2013almost,jiang2012joint}.
In fog computing, the presence of remote, heterogeneous devices on edge nodes motivated novel schemes
to match QoS requirements and maximize network usage. Authors of~\cite{brogi2017best}
focus on the provision of QoS constrained, eligible deployments for applications. The problem is showed NP-hard
with a reduction from the subgraph isomorphism problem. Preprocessing plus backtracking determines the
final eligible deployment restricting the search space. But, no performance target is optimized. 

In~\cite{Yu2018Prov}, application provisioning is studied from the perspective of the network infrastructure.
A fully polynomial-time approximation scheme is derived for single and  multiple application deployment, showing
large QoS performance improvement with respect to applications' bandwidth and delay figures; computational
requirements are not accounted for. 

Taneja et al.~\cite{taneja2017resource} define a placement algorithm by mapping the directed acyclic graph of the
modules of an IoT-based application into fog and cloud nodes. Numerical results show performance gains in terms
of latency, energy and bandwidth constraints compared to edge-agnostic placement schemes. In our work, conversely,
we provide an optimization framework to account for the coupling of traffic and computing demands of a batch of
applications to be deployed over multiple regions.


\section{Conclusions}


In this paper, we have introduced an optimization framework for microservice scheduling over 
fog infrastructures, where different configurations are used to orchestrate fog computation modules to the edge or in
cloud. The problem combines a multi-comodity flow and a placement problem, but can be reduced to a
$m$-dimensional knapsack problem by introducing throughput proportionality. We proposed a
greedy algorithm, namely FPA, which performs efficiently with respect to the optimal solution
by performing placement using a gradient approach. We have tested numerically our framework
under realistic dimensioning, leveraging our platform FogAtlas. Extensive numerical experiments
have confirmed the scalability properties of the proposed fog orchestration technique. 


\bibliographystyle{IEEEtran}
\bibliography{fog}

\end{document}